\documentclass[aps,showpacs,superscriptaddress,nofootinbib,preprint,11pt]{revtex4}
\usepackage{graphicx}
\usepackage{subfigure}
\def\slashchar#1{\setbox0=\hbox{$#1$}
   \dimen0=\wd0 \setbox1=\hbox{/} \dimen1=\wd1
   \ifdim\dimen0>\dimen1 \rlap{\hbox to \dimen0{\hfil/\hfil}} #1
   \else  \rlap{\hbox to \dimen1{\hfil$#1$\hfil}} / \fi}

\def\D{\slashchar{D}}

\begin{document}
\title{$\bar\nu$ induced $\bar K$ production off the nucleon}

\author{M. Rafi \surname{Alam}}
\affiliation{Department of Physics, Aligarh Muslim University, Aligarh-202 002, India}
\author{I. \surname{Ruiz Simo}}
\affiliation{Departamento de F\'\i sica Te\'orica and IFIC, Centro Mixto
Universidad de Valencia-CSIC, Institutos de Investigaci\'on de
Paterna, E-46071 Valencia, Spain}
\affiliation{Departamento de  F\'\i sica At\'omica  Molecular y Nuclear,
Universidad de Granada, E-18071 Granada, Spain}
\author{M. Sajjad \surname{Athar}}
\affiliation{Department of Physics, Aligarh Muslim University, Aligarh-202 002, India}
\author{M. J.  \surname{Vicente Vacas}}
\affiliation{Departamento de F\'\i sica Te\'orica and IFIC, Centro Mixto
Universidad de Valencia-CSIC, Institutos de Investigaci\'on de
Paterna, E-46071 Valencia, Spain}

\begin{abstract}
The charged current antikaon production off nucleons induced by antineutrinos  is studied at low and intermediate energies. We extend here our previous calculation on kaon production induced by neutrinos. We have developed 
a microscopic model  that starts from  the SU(3) chiral Lagrangians and includes background terms and the resonant mechanisms associated to the lowest lying resonance in the channel, namely,  the $\Sigma^*(1385)$.
Our results could be of interest for  the background  estimation of various neutrino oscillation experiments like MiniBooNE and SuperK. They can also be helpful for the planned $\bar \nu-$experiments 
like MINER$\nu$A, NO$\nu$A and T2K phase II and for  beta-beam experiments with antineutrino energies around 1~GeV.
\end{abstract}
\pacs{25.30.Pt,13.15.+g,12.15.-y,12.39.Fe}
\maketitle
\section{Introduction}
\label{Introduction}
Weak interaction experiments with neutrino energies around 1 GeV are quite sensitive to
the neutrino oscillation parameters and as a consequence many experiments like MiniBooNE,
SciBooNE, K2K, T2K, NO$\nu$A, etc. explore this energy range. Although many interesting results can be obtained without a detailed knowledge of the various processes used for the neutrino detection or the neutrino flux, 
a reliable estimate of the $\nu-$N cross section for various processes is mandatory to carry out a precise analysis of the measurements. 

Among these processes, strangeness conserving ($\Delta S=0$) weak interactions 
involving quasielastic production of leptons induced by charged as well as neutral weak currents have 
been widely studied~\cite{Boyd:2009zz,Leitner:2006ww,Leitner:2006sp,Benhar:2010nx,Martini:2010ex,Amaro:2010sd,Nieves:2011pp}. 
Much work has also  been done to understand one pion production in the weak sector~\cite{AlvarezRuso:1998hi,Sato:2003rq,Graczyk:2009qm,Hernandez:2007qq,Leitner:2008wx,Leitner:2010jv,Hernandez:2010bx,Lalakulich:2010ss}.
There are other inelastic reactions like hyperon and kaon production ($\Delta S=\pm 1$) that could also be measured even at quite low energies. However, very few calculations  study these processes~\cite{VicenteSingh,Mintz:2007zz,Dewan,Shrock,Amer:1977fy,Mart:2009,RafiAlam:2010kf}.  
This is partly justified by their small cross sections due to the Cabibbo suppression. 
As a result of this situation, the Monte Carlo generators used in the analysis of the current experiments apply models 
that are not well suited to describe the strangeness production at low energies.
 NEUT, for example, used by Super-Kamiokande, K2K, SciBooNE and 
T2K, only considers associated production of kaons within a model based on the excitation and later decay of baryonic resonances and from deep inelastic scattering (DIS)~\cite{Hayato:2009zz}. 
Similarly, other neutrino event generators like  NEUGEN~\cite{Gallagher:2002sf}, NUANCE~\cite{Casper:2002sd}
 (see also discussion in 
Ref.~\cite{Zeller:2003ey}) and GENIE~\cite{Andreopoulos:2009rq} do not consider single hyperon/kaon production.

Recently we have studied single kaon production induced by neutrinos at low and 
intermediate energies~\cite{RafiAlam:2010kf} using Chiral Perturbation Theory ($\chi$PT). 
We found that up to E$_{\nu_\mu}\approx 1.2$ GeV, 
single kaon production dominates over the associated production of kaons along with hyperons
which is mainly due to its lower threshold energy.

In this work, we extend our model to include weak single antikaon production off nucleons. 
 The theoretical model is necessarily more complicated 
than for kaons because resonant mechanisms, absent for the kaon case, could be relevant.  
On the other hand, the threshold for associated antikaon 
production corresponds to the $K-\bar K$ channel  and it is much higher than for the kaon case (KY). 
This implies that the process we study is the dominant source of antikaons for a wide range of energies.
 
The study may be useful in the analysis of antineutrino experiments at  MINER$\nu$A, 
NO$\nu$A, T2K and others. For instance, MINER$\nu$A has plans to investigate 
several strange particle production reactions with both neutrino and antineutrino beams~\cite{Solomey:2005rs} with high statistics. Furthermore,  the T2K experiment~\cite{Kobayashi:2005} as well as beta beam experiments~\cite{Mezzetto:2010} will work at energies where the single kaon/antikaon production may be important.

We introduce the formalism in Sec.~\ref{Formalism}. 
In Sec.~\ref{Results and Discussion}, we present the results, discussions and conclusions.

\section{Formalism}
\label{Formalism}
\begin{figure}
\begin{center}
\includegraphics[width=0.8\textwidth,height=.4\textwidth]{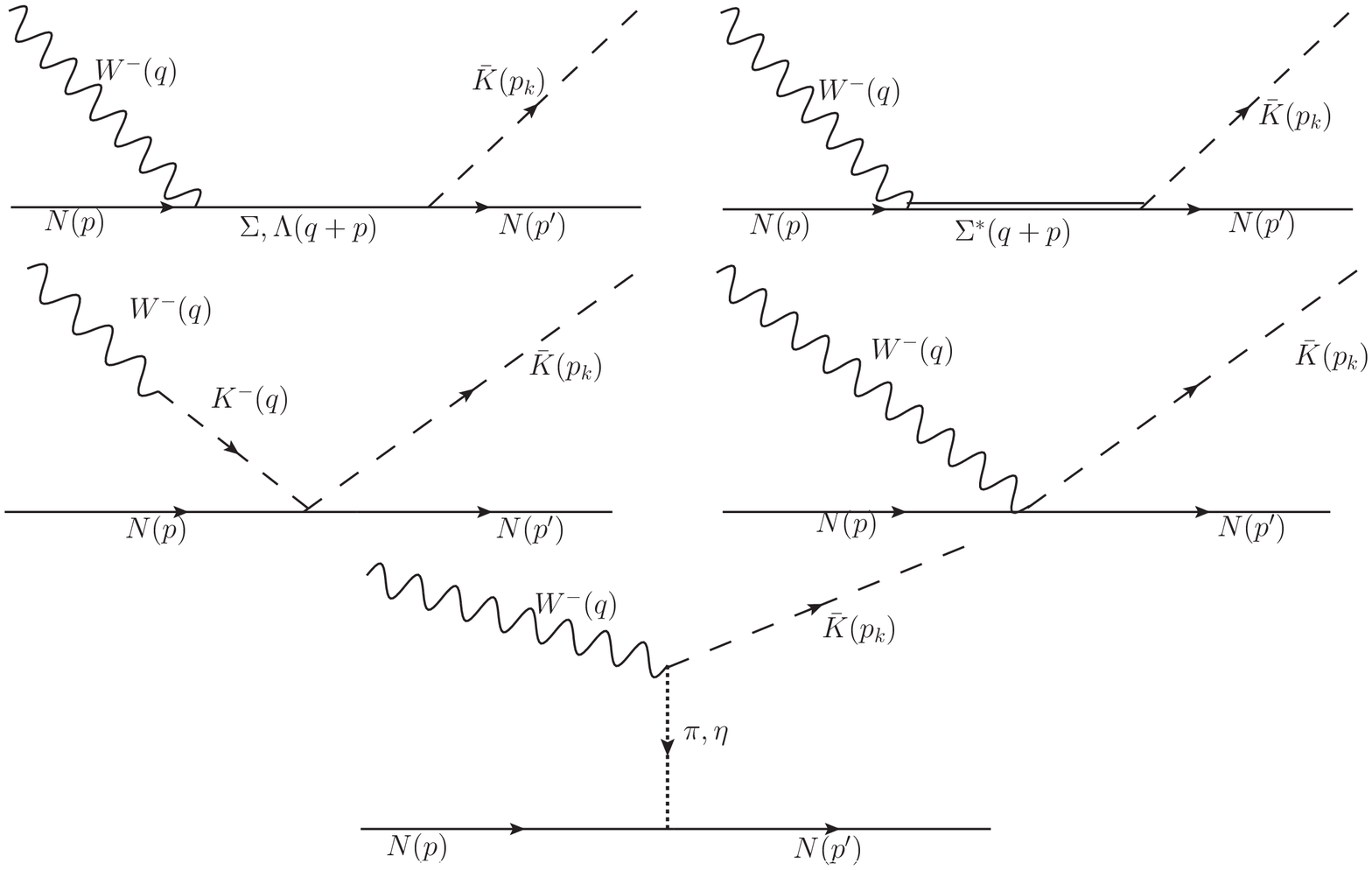}
\caption{Feynman diagrams for the process $\bar \nu N\rightarrow l N^\prime \bar K$.  First row 
from left to right:  s-channel $\Sigma,\Lambda $ propagator (labeled SC in the text), s-channel $\Sigma^*$
 Resonance (SCR), second row: kaon pole term (KP); Contact term (CT)
and last row: Pion(Eta) in flight ($ \pi P/ \eta P $). }
\label{fg:terms}
\end{center}
\end{figure}

The basic reaction for antineutrino induced charged current  antikaon production  is
\begin{equation}\label{reaction}
\bar \nu_{l}(k) + N(p) \rightarrow l(k^{\prime}) + N^\prime(p^{\prime}) + \bar K(p_{k}) ,
\end{equation}
where $l=e^+,\mu^+$ and $ N \& N^\prime $ are nucleons.
 The expression for the differential cross section in the laboratory frame for the above process is given by
\begin{eqnarray}\label{sigma_inelas}
d^{9}\sigma &=& \frac{1}{4 M E(2\pi)^{5}} \frac{d{\vec k}^{\prime}}{ (2 E_{l})} 
\frac{d{\vec p\,}^{\prime}}{ (2 E^{\prime}_{p})} \frac{d{\vec p}_{k}}{ (2 E_{k})}
 \delta^{4}(k+p-k^{\prime}-p^{\prime}-p_{k})\bar\Sigma\Sigma | \mathcal M |^2,
\end{eqnarray}
where  $ k( k^\prime) $ is the momentum of the incoming(outgoing) lepton with energy $E( E^\prime)$,  $p( p^\prime)$ is the momentum of the incoming(outgoing)
nucleon. The kaon 3-momentum is $\vec{p}_k $ having energy $ E_k  $, $M$ is the nucleon mass,
$ \bar\Sigma\Sigma | \mathcal M |^2  $ is the square of the transition amplitude
 averaged(summed) over the spins of the initial(final) state. 
It can be written as 
\begin{equation}
\label{eq:Gg}
 \mathcal M = \frac{G_F}{\sqrt{2}}\, j_\mu J^{\mu}=\frac{g}{2\sqrt{2}}j_\mu \frac{1}{M_W^2}
\frac{g}{2\sqrt{2}}J^{\mu},
\end{equation}
 where $j_\mu$ and $  J^{\mu}$ are the leptonic and hadronic currents respectively, 
$G_F=\sqrt{2} \frac{g^2}{8 M^2_W}$ is the Fermi coupling constant, 
$g$ is the gauge coupling and $M_W$ is the mass of the $W$-boson.
The leptonic current can be readily obtained from the standard model  Lagrangian coupling the $W$ bosons to the leptons 
\begin{equation}
{\cal L}=-\frac{g}{2\sqrt{2}}\left[j^\mu{ W}^-_\mu+h.c.\right].
\end{equation}
We construct a model including non resonant terms and the decuplet resonances,
that couple strongly to the pseudoscalar mesons. 
The same approach successfully describes  the pion production case 
(see for example  Ref.~\cite{Hernandez:2007qq}).
The channels that  contribute to the hadronic current are depicted in Fig.~\ref{fg:terms}. 
There are s-channels with  $\Sigma,\Lambda$(SC) and $\Sigma^*$(SCR) as intermediate states, a kaon pole (KP) term, 
a contact term (CT),
 and finally a meson ($\pi$P,$\eta$P) exchange term. For these specific reactions 
there are no u-channel processes with hyperons in the intermediate state. 

The contribution coming from different terms can be obtained from the $\chi$PT Lagrangian.
We follow the conventions of Ref.~\cite{Scherer:2002tk} to write the
 lowest-order SU(3) chiral Lagrangian describing the interaction of pseudoscalar mesons in the 
presence of an external current,
\begin{equation}
\label{eq:lagM}
{\cal L}_M^{(2)}=\frac{f_\pi^2}{4}\mbox{Tr}[D_\mu U (D^\mu U)^\dagger]
+\frac{f_\pi^2}{4}\mbox{Tr}(\chi U^\dagger + U\chi^\dagger),
\end{equation}
where the parameter $f_\pi=92.4$ MeV is the pion  decay constant, $U(x)=\exp\left(i\frac{\phi(x)}{f_\pi}\right)$ 
is the SU(3) representation of the meson fields $\phi(x)$
and $D_\mu U$ is its covariant derivative
\begin{eqnarray}
D_\mu U&\equiv&\partial_\mu U -i r_\mu U+iU l_\mu\,.
\end{eqnarray}
For the charged current case the left and right handed currents $l_\mu$ and $r_\mu$ are given by
\begin{equation}
r_\mu=0,\quad l_\mu=-\frac{g}{\sqrt{2}}
({W}^+_\mu T_+ + {W}^-_\mu T_-),
\end{equation}
with $W^\pm$  the $W$ boson fields and
$$
T_+=\left(\begin{array}{rrr}0&V_{ud}&V_{us}\\0&0&0\\0&0&0\end{array}\right);\quad
T_-=\left(\begin{array}{rrr}0&0&0\\V_{ud}&0&0\\V_{us}&0&0\end{array}\right).
$$
Here,  $V_{ij}$ are the elements of the 
Cabibbo-Kobayashi-Maskawa  matrix.
The second term of the Lagrangian of Eq.~\ref{eq:lagM}, that incorporates the
explicit breaking of chiral symmetry coming from the quark masses~\cite{Scherer:2002tk}, is not relevant for our study.

The lowest-order chiral Lagrangian describing the interaction between baryon-meson
octet in the presence of an external weak current 
can be written in terms of the SU(3) matrix as
\begin{equation}
\label{eq:lagB}
{\cal L}^{(1)}_{MB}=\mbox{Tr}\left[\bar{B}\left(i\D
-M\right)B\right]
-\frac{D}{2}\mbox{Tr}\left(\bar{B}\gamma^\mu\gamma_5\{u_\mu,B\}\right)
-\frac{F}{2}\mbox{Tr}\left(\bar{B}\gamma^\mu\gamma_5[u_\mu,B]\right),
\end{equation}
where $M$ denotes the mass of the baryon octet, and the parameters $D=0.804$ and $F=0.463$
can be determined from the baryon semileptonic decays~\cite{Cabibbo:2003cu}.
The covariant derivative of $B$ is given by
\begin{equation}
D_\mu B=\partial_\mu B +[\Gamma_\mu,B],
\end{equation}
with
\begin{equation}
\Gamma_\mu=\frac{1}{2}\left[u^\dagger(\partial_\mu-ir_\mu)u
+u(\partial_\mu-il_\mu)u^\dagger\right],
\end{equation}
where  we have introduced $u^2=U$. Finally, 
\begin{equation}
\label{Eq:octet-weak_int}
u_\mu= i\left[u^\dagger(\partial_\mu-i r_\mu)u-u(\partial_\mu-i
l_\mu)u^\dagger\right].
\end{equation}
The next order meson baryon Lagrangian contains many new terms (see for instance Ref.~\cite{Oller:2006yh}). 
Their importance for kaon production will be small at low energies and there are some uncertainties 
in the coupling constants. Nonetheless, for consistency with previous calculations, 
we will include the contribution to the weak magnetism coming from the pieces 
\begin{equation}
{\cal L}^{(2)}_{MB}= d_5 \mbox{Tr}\left(\bar{B}[f_{\mu\nu}^+,\sigma^{\mu\nu}B]\right)+
d_4 \mbox{Tr}\left(\bar{B}\{f_{\mu\nu}^+,\sigma^{\mu\nu}B\}\right)+\dots,
\end{equation}
where the tensor $f_{\mu\nu}^+$ can be reduced for our study to
\begin{equation}
 f_{\mu\nu}^+=\partial_\mu l_\nu-\partial_\nu l_\mu -i [l_\mu,l_\nu].
\end{equation}
In this case, the coupling constants are fully determined by the proton and neutron anomalous magnetic moments.
The same approximation has also been used in calculations of single pion~\cite{Hernandez:2007qq} and  kaon production
~\cite{RafiAlam:2010kf} induced by neutrinos.

As it is the case for the $\Delta(1232)$ in pion production, we expect that the weak excitation of the 
$\Sigma^*(1385)$ resonance and its subsequent decay in $NK$ may be important. 
The lowest order SU(3) Lagrangian coupling the pseudoscalar mesons
with decuplet-octet baryons in presence of
external weak current is given by
\begin{equation}
{\cal L}_{dec} = {\cal C} \left( \epsilon^{abc} 
\bar T^\mu_{ade} u_{\mu,b}^d B_c^e + 
\, h.c. \right),
 \label{eq:dec_lag}
\end{equation}
where $T^\mu$ is the SU(3) representation of the decuplet fields, $a-e$ are flavour 
indices\footnote{The physical states of the decuplet are: $T_{111} = \Delta^{++},T_{112} = \frac{\Delta^{+}}{\sqrt3},
T_{122} = \frac{\Delta^{0}}{\sqrt3}, T_{222} = \Delta^{-},T_{113} = \frac{\Sigma^{*+}}{\sqrt3},
T_{123} = \frac{\Sigma^{*0}}{\sqrt6},T_{223} = \frac{\Sigma^{*-}}{\sqrt3},T_{113} =  \frac{\Xi^{+}}{\sqrt3},
T_{133} =  \frac{\Xi^{0}}{\sqrt3},T_{333} = \Omega^{-} $.},
$B$ corresponds to the baryon octet and $u_\mu$ is the SU(3) representation 
of the pseudoscalar mesons interacting with weak left $l_\mu$ and right $r_\mu$ handed currents 
(See Eq.~\ref{Eq:octet-weak_int}). The parameter
${\cal C}\simeq 1$  has been fitted to the $\Delta(1232)$  decay-width.  
The spin 3/2  propagator for
$\Sigma^*$  is given by
\begin{equation}
G^{\mu\nu}(P)= \frac{P^{\mu\nu}_{RS}(P)}{P^2-M_{\Sigma^*}^2+ i M_{\Sigma^*} \Gamma_{\Sigma^*}},
\qquad 
\end{equation}
where $P=p+q$ is the momentum carried by the resonance, $q=k-k^\prime$ and 
$P^{\mu \nu}_{RS}$ is the projection operator
\begin{equation}
P^{\mu\nu}_{RS}(P)= \sum_{spins} \psi^{\mu} \bar \psi^{\nu} =- (\slashchar{P} + M_{\Sigma^*}) \left [ g^{\mu\nu}-
  \frac13 \gamma^\mu\gamma^\nu-\frac23\frac{P^\mu
  P^\nu}{M_{\Sigma^*}^2}+ \frac13\frac{P^\mu
  \gamma^\nu-P^\nu \gamma^\mu }{M_{\Sigma^*}}\right],
\label{eq:rarita_prop}
\end{equation}
with $M_{\Sigma^*}$ the resonance mass  and $\psi^{\mu}$ 
 the Rarita-Schwinger spinor. The  $\Sigma^*$ width obtained using the
Lagrangian  of Eq.~\ref{eq:dec_lag}  can be written as
\begin{eqnarray}
 \Gamma_{\Sigma^*}&=&\Gamma_{\Sigma^*\rightarrow \Lambda \pi} 
+ \Gamma_{\Sigma^*\rightarrow \Sigma \pi}+ \Gamma_{\Sigma^*\rightarrow N \bar{K}}\; ,
\label{eq:width}
\end{eqnarray}
where
\begin{eqnarray}
 \Gamma_{\Sigma^* \rightarrow Y,\, meson}&=&\frac{C_Y}{192\pi}\left(\frac{\cal C}{f_\pi}\right)^2
\frac{(W+M_Y)^2-m^2}{W^5}\lambda^{3/2}(W^2,M_Y^2,m^2) \,
\Theta(W-M_Y-m).
\end{eqnarray}
Here, $m,\, M_Y$ are the masses of the emitted meson and baryon.
 $\lambda(x,y,z)=(x-y-z)^2-4yz$ and $\Theta$ is the 
step function. The factor $C_Y$   is 1 for $\Lambda$ and $\frac23$ for $N $ and $\Sigma$.

Using symmetry arguments, the most general $W^- N \rightarrow \Sigma^*$ vertex can be written in terms of 
a vector and an axial-vector part as,
\begin{eqnarray} \label{eq:delta_amp}
\langle \Sigma^{*}; P= p+q\, | V^\mu | N;
p \rangle &=& V_{us} \bar\psi_\alpha(\vec{P} ) \Gamma^{\alpha\mu}_V \left(p,q \right)
u(\vec{p}\,), \nonumber \\
\langle \Sigma^{*}; P= p+q\, | A^\mu | N;
p \rangle &=& V_{us} \bar \psi_\alpha(\vec{P} ) \Gamma^{\alpha\mu}_A \left(p,q \right)
u(\vec{p}\,),
\end{eqnarray}
where
\begin{eqnarray}
\Gamma^{\alpha\mu}_V (p,q) &=&
\left [ \frac{C_3^V}{M}\left(g^{\alpha\mu} \slashchar{q}-
q^\alpha\gamma^\mu\right) + \frac{C_4^V}{M^2} \left(g^{\alpha\mu}
q\cdot P- q^\alpha P^\mu\right)
+ \frac{C_5^V}{M^2} \left(g^{\alpha\mu}
q\cdot p- q^\alpha p^\mu\right) + C_6^V g^{\mu\alpha}
\right ]\gamma_5 \nonumber\\
\Gamma^{\alpha\mu}_A (p,q) &=& \left [ \frac{C_3^A}{M}\left(g^{\alpha\mu} \slashchar{q}-
q^\alpha\gamma^\mu\right) + \frac{C^A_4}{M^2} \left(g^{\alpha\mu}
q\cdot P- q^\alpha P^\mu\right)
+ C_5^A g^{\alpha\mu} + \frac{C_6^A}{M^2} q^\mu q^\alpha
\right ]. \label{eq:del_ffs}
 \end{eqnarray}
Our knowledge of these form factors is quite limited. The Lagrangian of Eq.~\ref{eq:dec_lag} gives us only $C_5^A(0)=-2{\cal{C}}/\sqrt{3}$ (for the $\Sigma^{*-}(1385)$ case).
However, using SU(3) symmetry we can relate all other form factors to those of the $\Delta(1232)$ resonance, such that
$C_i^{\Sigma*^-}/C_i^{\Delta^{+}}=-1$ and $C_i^{\Sigma*^-}/{C_i^{\Sigma*^0}}=\sqrt{2}$.  See Refs.~\cite{AlvarezRuso:1998hi,Paschos:2005,Hernandez:2007qq,Leitner:2008ue,Hernandez:2010bx} for details of the $W N\Delta$ form-factors.
In the $\Delta$ case, the vector form factors are relatively well known from electromagnetic processes and there is some information on the axial ones from the study of pion production.
We will use the same set as in Ref.~\cite{Hernandez:2007qq,Hernandez:2010bx}, where pion production induced by neutrinos has been studied, except for $C_5^A(0)$, obtained directly from the Lagrangian and $C_6^A$.
These latter two form factors are related by PCAC so that $C_6^A=C_5^A M^2/(m_K^2-q^2)$.

In our model, we use an SU(3) symmetric Lagrangian. The only SU(3) breaking comes from the use of
physical masses. This is expected to be a good description for the background terms, as it was discussed for the kaon production induced by neutrinos in Ref.~\cite{RafiAlam:2010kf}. Little is known about the SU(3) breaking for the axial couplings of the baryon decuplet, but only a small breaking has been found for their electromagnetic properties~\cite{hep-ph/9211247,arXiv:0907.0631}. Therefore, we can expect a similarly small uncertainty in the size of the $\Sigma^*(1385)$ contribution.

 Even from relatively low neutrino energies,
other baryonic resonances, beyond the $\Sigma^*(1385)$, could contribute to the cross section, as they are  close to the kaon nucleon threshold. However, their weak couplings are basically unknown. Also, the theoretical estimations 
of these couplings are still quite uncertain. Nonetheless, recent advances on the radiative decays of these resonances, both experimental and theoretical (see, e.g., Refs.~\cite{Doring:2006ub,Taylor:2005zw}) are very promising and may help to develop a more complete model in the future.

Finally, we consider  the $q^2$ dependence of the weak current couplings provided by the chiral Lagrangians.  In this work, we follow the same procedure as in Ref.~\cite{RafiAlam:2010kf}\footnote{A more elaborate discussion can be found there.}
and adopt a global dipole form factor
$
F(q^2)=1/(1-q^2/M_F^2)^2,
$ 
with a mass $M_F\simeq 1$ GeV that multiplies all the hadronic currents, except the resonant one,
 that has been previously  discussed.  Its effect, that should be small at low neutrino energies, will give an idea of the uncertainties of the calculation and will be explored in the next section.

Detailed expressions of the resulting hadronic currents   $J^{\mu}$  containing both background and resonant terms are listed in the appendix \ref{app:amplitude}.

\section{Results and Discussion}
\label{Results and Discussion}

We consider the following strangeness changing ($| \Delta S | = 1$) charged-current reactions:
\begin{eqnarray}\label{processes}
\bar \nu_l + p &\rightarrow & l^+ + K^- + p  \nonumber\\
\bar \nu_l + p &\rightarrow & l^+ +\bar K^0 + n  \nonumber\\
\bar \nu_l + n &\rightarrow & l^+ + K^- + n \,.
\end{eqnarray}

\begin{figure}[htb]
\begin{center}
\includegraphics[width=0.6\textwidth]{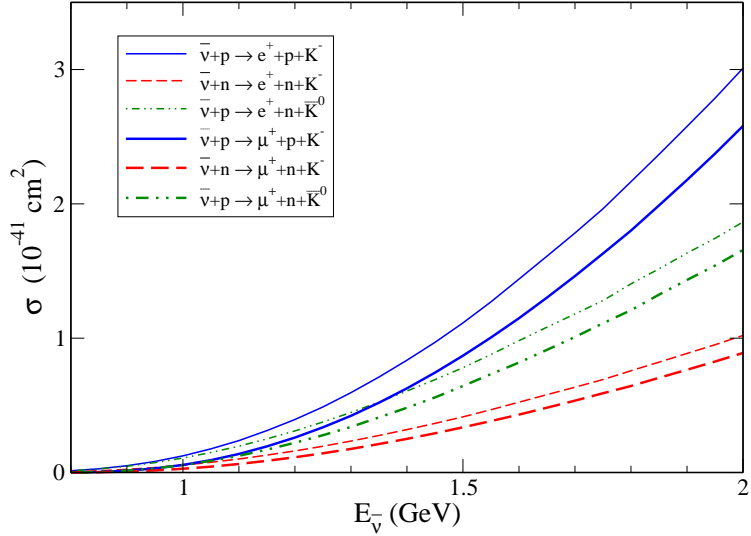}
\caption{Cross-section for the processes 
$\bar\nu_\mu  N\rightarrow \mu^+ N^\prime \bar K$  and
$\bar\nu_e    N\rightarrow e^+ N^\prime \bar K$
as a function of the antineutrino energy}
\label{fg:xsec_all_chnl}
\end{center}
\end{figure}
In Fig.~\ref{fg:xsec_all_chnl}, we show their total cross section for  electronic and muonic antineutrinos as a function of energy. We obtain similar values to the  cross sections  of kaon production  induced by neutrinos of Ref.~\cite{RafiAlam:2010kf},  even when there are no resonant contributions.
The electronic antineutrino cross sections are slightly larger, but they do not present any other distinguishing feature. For all channels, the cross sections are very small, as compared to other processes induced by antineutrinos at these energies, like pion production, due to the Cabibbo suppression and to the smallness of the available phase space.
 Nonetheless, the reactions we have studied are the main source of antikaons for a wide range of neutrino energies.
In fact, the lowest energy antikaon associate production, ($K\bar{K}$, $| \Delta S | = 0$), has a quite high threshold ($\approx 1.75$ GeV) and thus, it leads to even smaller cross sections in the range of energies we have explored.
For instance, at 2 GeV, GENIE predicts antikaon production cross sections at least two orders of magnitude smaller than our calculation\footnote{This has been obtained with GENIE version 2.7.1 and corresponds to $K\bar{K}$ processes. }.

As it was expected, our results would lead to a very minor signal in past experiments. 
For instance, we  have evaluated the flux averaged cross-section $\langle\sigma\rangle$ for the MiniBooNE 
antineutrino flux~\cite{AguilarArevalo:2011sz} in the sub GeV energy region. The results are given in Table~\ref{tb:flux} and compared with the recent measurement of the neutral current $\pi^0$ production per nucleon with the same flux~\cite{AguilarArevalo:2009ww}.
\begin{table*}[htb]
\begin{center}
\caption{$\langle\sigma\rangle$  for $\bar K$ production with MiniBooNE $\bar\nu_\mu$ flux and
neutral current $\pi^0$ production (per nucleon) measured at MiniBooNE~\cite{AguilarArevalo:2009ww}}. 
\begin{tabular}{| l |c |}\hline\hline
Process & $\langle\sigma\rangle$ ($10^{-41}$ cm$^2$)  \\ \hline
$\bar \nu_\mu + p \rightarrow  \mu^+ + K^- + p$ & 0.11  \\
$\bar \nu_\mu + p \rightarrow \mu^+ +\bar K^0 + n $& 0.08 \\
$\bar \nu_\mu + n \rightarrow  \mu^+ + K^- + n$& 0.04  \\ \hline
$\bar \nu_\mu + ^{12}C \rightarrow \bar \nu_\mu  +X + \pi^0 $   & $14.8\pm 0.5\pm 2.3$\\ \hline
\end{tabular}
\label{tb:flux}
\end{center}
\end{table*}
We find that the antikaon production cross section is  around two orders of magnitude smaller than the  NC $\pi^0$ one  at MiniBooNE.  Given the number of neutral pions observed for the antineutrino beam we expect that only a few tens of antikaons were produced in this experiment. One should notice here that the average antineutrino energy at MiniBooNE is well below the kaon threshold. Thus, we are only sensitive to the high energy tail of the flux.

One could expect a relatively larger signal for  the atmospheric neutrino
$\bar\nu_e$ and $\bar\nu_\mu$ induced events at SuperK, given the larger neutrino energies. But even there we find a very small background from antikaon events.
Taking the antineutrino fluxes from Ref.~\cite{Honda:2006qj} we have calculated the event rates for the 22.5kT water target and a period of 1489 days as in the SuperK analysis of Ref.~\cite{Ashie:2005ik}. We obtain 0.8
 $e^+$ and 1.5 $\mu^+$ events. Although the model has large uncertainties at
high energies, the rapid fall of the neutrino spectrum implies that the high
energy tail contributes very little to the background.

 We have also estimated the average cross sections for the expected antineutrino fluxes at
T2K~\cite{arXiv:1109.3262} and MINER$\nu$A (low energy configuration)~\cite{minflux}. In both cases, we have implemented an energy cut ($E_k+E_l<2$ GeV), that insures that high energy neutrinos, for which our model is less reliable, play a minor role. The results are presented in Table~\ref{tb:fluxes}. For T2K, we get similar results to the MiniBooNE case whereas the average cross section is much larger at  MINER$\nu$A because of the higher neutrino energies.
 \begin{table*}[htb]
\begin{center}
\caption{$\langle\sigma\rangle$ ($10^{-41}$ cm$^2$) for $\bar K$ production with $\bar\nu_\mu$ T2K~\cite{arXiv:1109.3262} and MINER$\nu$A~\cite{minflux} expected fluxes.} 
\begin{tabular}{| l |c |c|}\hline\hline
Process & $\langle\sigma\rangle$ MINER$\nu$A &  $\langle\sigma\rangle$ T2K \\ \hline
$\bar \nu_\mu + p \rightarrow  \mu^+ + K^- + p$    & 1.1 & 0.07\\
$\bar \nu_\mu + p \rightarrow \mu^+ +\bar K^0 + n $& 0.49& 0.04\\
$\bar \nu_\mu + n \rightarrow  \mu^+ + K^- + n$    & 0.33& 0.02\\  \hline
\end{tabular}
\label{tb:fluxes}
\end{center}
\end{table*}

Hitherto, our results correspond to relatively low antineutrino energies, where
our model is best suited. However,  the model could also be used
to compare with data obtained at much higher neutrino energies selecting events such that the invariant mass of 
hadronic part is close to antikaon-nucleon threshold and the transferred momentum $q$ is small.
This  procedure has been used, for instance, in the analysis of two pion production induced by neutrinos~\cite{Adjei:1981nw,Kitagaki:1986ct}.

\begin{figure}[htb]
\begin{center}
\includegraphics[width=0.6\textwidth]{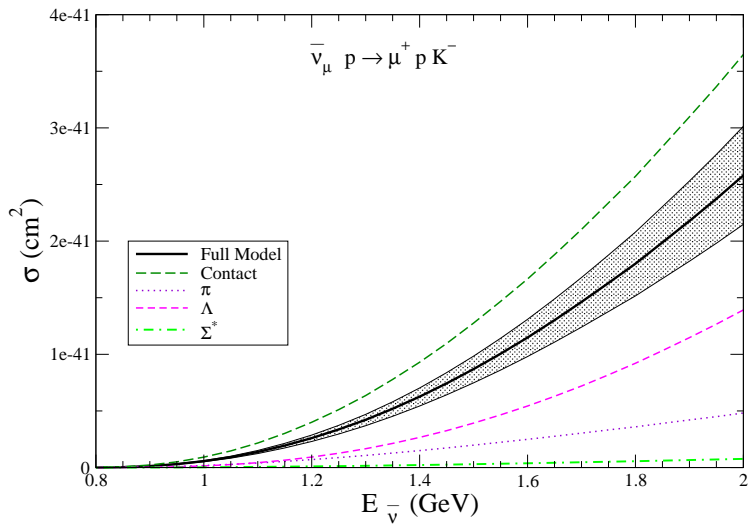}
\caption{Cross-section for the process $\bar \nu_\mu p \rightarrow \mu^+ p K^- $.}
\label{fg:xsec_mu_pp}
\end{center}
\end{figure}

In Fig.~\ref{fg:xsec_mu_pp}, we show the size of several contributions to the  $\bar \nu_\mu p \rightarrow \mu^+ p K^- $ reaction. 
Obviously, this separation is not an observable and only the full cross section obtained with the sum of the amplitudes has a physical sense. However, it could help us to get some idea of how the uncertainties associated to some of the mechanisms, like the $\Sigma^*(1385)$ one, could affect our results.
The cross section is clearly dominated by the non--resonant terms, providing the CT term the largest contribution. We see the destructive interference that leads to a total cross section smaller than that predicted by the CT term alone. We could also remark the negligible contribution of the $\Sigma^*(1385)$ channel. This fact is at variance with the strong $\Delta$ dominance for pion production and it can be easily understood because the  $\Sigma^*$ mass is below the kaon production threshold.
We have also explored, the uncertainties associated with the form factor.
The curve labeled as ``Full Model'' has been calculated with a dipole form factor with a mass of 1 GeV. The band corresponds to  a 10 percent variation of this parameter. The effect is similar in the other channels and we will only show the results for the central value of 1 GeV. 
\begin{figure}[htb]
\begin{center}
\includegraphics[width=0.6\textwidth]{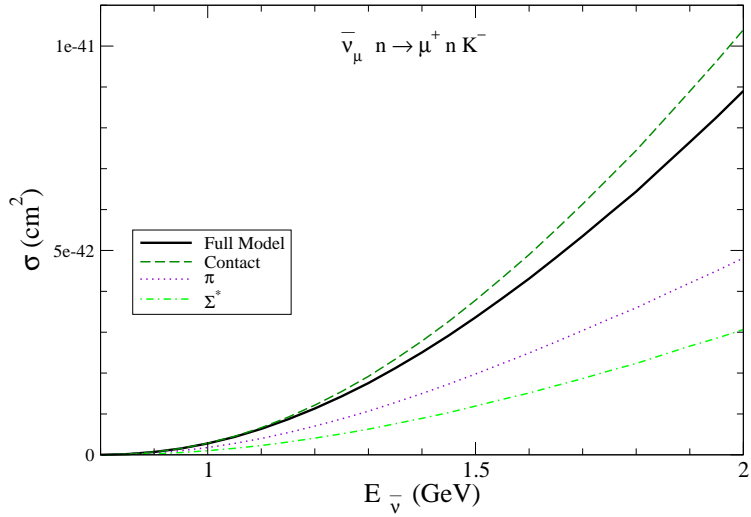}
\caption{Cross-section for $\bar \nu_\mu n \rightarrow \mu^+ n K^- $.} 
\label{fg:xsec_mu_nn}
\end{center}
\end{figure}
\begin{figure}[htb]
\begin{center}
\includegraphics[width=0.6\textwidth]{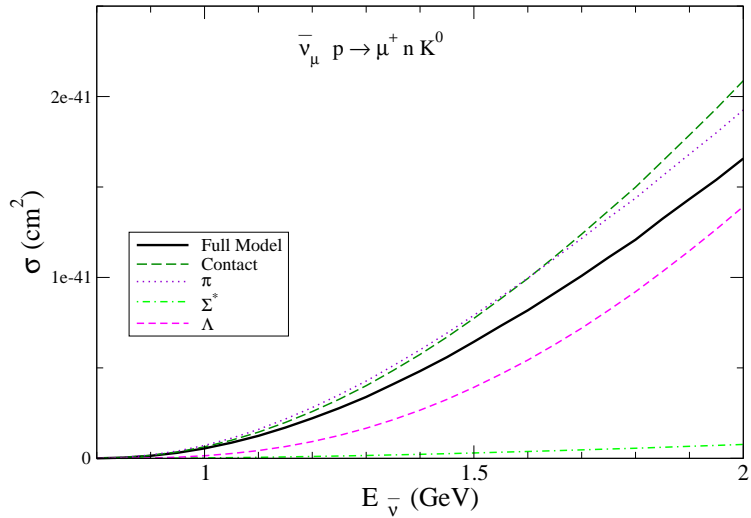}
\caption{Cross-section for $\bar \nu_\mu p \rightarrow \mu^+ n \bar{K}^0 $.} 
\label{fg:xsec_mu_pn}
\end{center}
\end{figure}
In Figs.~\ref{fg:xsec_mu_nn} and \ref{fg:xsec_mu_pn}, we show the other two channels. As in the previous case the CT term is very important. We observe, however, that the pion-pole term gives a contribution  as large as the CT one for the $\bar \nu_\mu p \rightarrow \mu^+ n K^0 $ process. For the  $\bar \nu_\mu n \rightarrow \mu^+ n K^- $ case, we find a substantial contribution of the $\Sigma^*$ resonance, due to the larger value of the couplings (see Table~\ref{tb:currents}). As in the first case, there is some destructive interference between the different mechanisms participating in these processes.

\begin{figure}[htb]
\begin{center}
\includegraphics[width=0.6\textwidth]{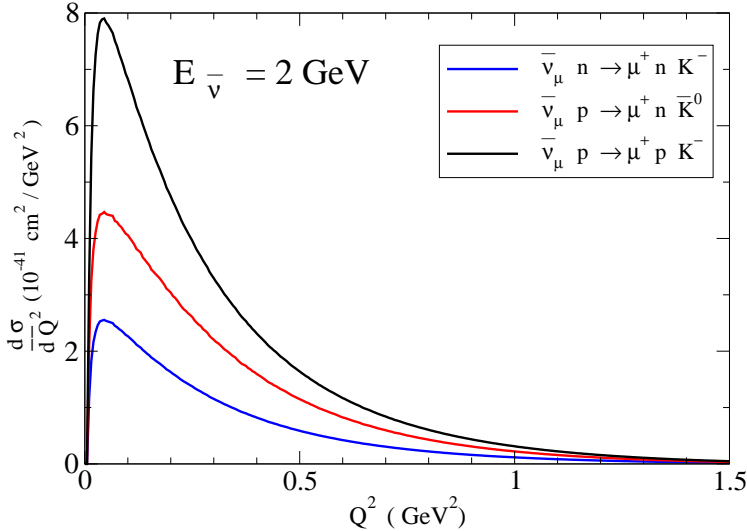}
\caption{$d\sigma/dQ^2$ cross section.}
\label{fg:q2}
\end{center}
\end{figure}
 In Fig.~\ref{fg:q2}, we show the $Q^2$ distributions for the three channels at a
antineutrino energy $E_{\bar{\nu}} = 2$ GeV. We have checked that the reactions are always forward peaked (for the final lepton),even in the absence of any form factor, favouring relatively small values of the
momentum transfer. We should notice however, that the smallness of $Q^2$ does not imply that $q^0$ or $\vec{q}$ are also small. In fact, because of the kaon mass both energy and momentum transfer are always large. Also nucleon laboratory momentum, even at threshold, is quite large ($\sim$ 0.48 GeV). This implies that, for these processes, 
Pauli blocking in nuclei would be  ineffective.

In summary, we have developed a microscopical model for single antikaon production off
nucleons induced by neutrinos based on the SU(3) chiral Lagrangians, including the lowest lying octet and decuplet baryons. This model is an extension of that of Ref.~\cite{RafiAlam:2010kf}, where single kaon production was investigated. The calculation is necessarily more complex for antikaons because resonant mechanisms, absent for the kaon case, could be relevant.  On the other hand, the threshold for associated antikaon 
production corresponds to the $K-\bar K$ channel  and it is much higher than for the kaon case (kaon-hyperon). 
This implies that the process we study is the dominant source of antikaons for a wide range of energies.
All parameters of the model involving only octet baryons are well known: Cabibbo's angle, $f_\pi$, the pion decay constant, the proton and neutron magnetic moments and the axial vector coupling constants  D and F.
The weak couplings of the $\Sigma^*(1385)$ have been obtained from those of the $\Delta(1232)$ using SU(3) symmetry.
Although they contain considerable uncertainties, we find that the resonance contribution is quite small.

We obtain for the single antikaon production cross sections similar to those of single kaon production, and around two orders of magnitude smaller that for pion production for antineutrino fluxes such as that from MiniBooNE. Nonetheless,
the study may be useful in the analysis of antineutrino experiments at  MINER$\nu$A, 
NO$\nu$A, T2K and others with high statistics and/or higher antineutrino energies.

\begin{acknowledgments}
This work is partly supported by DGICYT Contracts No. FIS2006-03438 and FIS2008-01143, 
the Generalitat Valenciana in the program Prometeo
and the EU Integrated Infrastructure Initiative Hadron
Physics Project under contract RII3-CT-2004-506078. I.R.S. acknowledges support from
the Ministerio de Educaci\'on. M.R.A. wishes to acknowledge the financial support
from the University of Valencia and Aligarh Muslim University under the academic exchange program 
and also to the DST, Government of India for the financial support
under the grant SR/S2/HEP-0001/2008.
\end{acknowledgments}

\appendix{\bf APPENDIX}

\section{Hadronic Currents}\label{app:amplitude}
For consistency with Eq. \ref{eq:Gg} the 
contributions to the hadronic current are 
\begin{eqnarray*}
J^\mu \arrowvert_{CT} &=&i A_{CT} V_{us} \frac{ \sqrt{2}}{2 f_\pi}  \bar N(p^\prime) \; (\gamma^\mu + B_{CT} \; \gamma^\mu \gamma_5 ) \; N(p) \\
J^\mu \arrowvert_{\Sigma} &=&i A_{\Sigma} (D-F) V_{us} \frac{ \sqrt{2}}{2 f_\pi} \bar N(p^\prime) p_k\hspace{-.9em}/ \; \gamma_5  \frac{ p\hspace{-.5em}/ +
  q\hspace{-.5em}/ + M_\Sigma}
  {( p +  q)^2 -M_\Sigma^2} \left(\gamma^\mu +i  \frac{(\mu_p + 2\mu_n)}{2 M} \sigma^{\mu \nu} q_\nu \right. \\
&+& \left. (D-F) \left\{ \gamma^\mu 
  - \frac{q^\mu}{ q^2-{M_k}^2 } q\hspace{-.5em}/ \right\} \gamma^5 \right) N(p) \\
J^\mu \arrowvert_{\Lambda} &=&  i A_{\Lambda} V_{us} (D+3F)  \frac{1} {2 \sqrt{2} f_\pi}   \bar N(p^\prime) p_k\hspace{-.9em}/ \; \gamma^5 \frac{ p\hspace{-.5em}/ +
  q\hspace{-.5em}/ +M_\Lambda}
  {( p +  q)^2 -M_\Lambda^2} \left(\gamma^\mu +i \frac{\mu_p}{2 M}  \sigma^{\mu \nu} q_\nu \right. \\
 &-& \left. \frac{(D + 3 F)}{3} \left\lbrace \gamma^\mu   - \frac{q^\mu }{ q^2-{M_k}^2 } 
q\hspace{-.5em}/ \right\rbrace \gamma^5 \right) N(p) \\
J^\mu \arrowvert_{KP}&=& i A_{KP} V_{us} \frac{\sqrt{2}}{2 f_\pi}  \bar N(p^\prime)  q\hspace{-.5em}/ \; N(p) \frac{q^\mu}{q^2-M_k^2}    \\
J^\mu \arrowvert_{\pi} &=& iA_{\pi } \frac{M\sqrt{2}}{2 f_\pi}  V_{us}  (D + F)\frac{ 2 {p_k}^\mu -q^\mu}{(q-p_k)^2 - {m_\pi}^2} \bar N(p^\prime)  \gamma_5  N(p) \\
J^\mu \arrowvert_{\eta} &=&i A_{\eta } \frac{M\sqrt{2}}{2 f_\pi}  V_{us}  (D - 3 F)\frac{2 {p_k}^\mu - q^\mu}{(q-p_k)^2 - {m_\eta}^2} \bar N(p^\prime) 
        \gamma_5 N(p) \\
J^\mu \arrowvert_{\Sigma^*} &=&- i A_{\Sigma^*} \frac{\cal C}{ f_\pi } \frac{1}{\sqrt{6}} \; V_{us} \;  
  \frac{p_k^\lambda}{P^2 - M_{\Sigma^*}^2 + i \Gamma_{\Sigma^*} M_{\Sigma^*}}\; 
\bar N(p^\prime) P_{RS_{\lambda \rho}} ( \Gamma_V^{\rho \mu} +\Gamma_A^{\rho \mu} ) N(p) 
\end{eqnarray*}
In $\Gamma_V^{\rho \mu} +\Gamma_A^{\rho \mu}$, the form factors are taken as for the $\Delta^+$ case. The extra factors for each of the $\Sigma^*$ channels are given by $A_{\Sigma^*} $ in Tab. \ref{tb:currents}.

\begin{table*}
 \begin{center}
\begin{tabular}{|l|c|c|c|c|c|c|c|c|} \hline \hline
Process 	& $B_{CT}$  & $A_{CT}$ & $A_{\Sigma}$ & $A_{\Lambda}$ & $A_{KP}$ & $A_{\pi }$ & $A_{\eta }$ & $ A_{\Sigma^*} $ \\\hline
$ \bar \nu   n \rightarrow l^+  K^-  n $     &    D-F  &  1 & -1 & 0 & -1 & 1 & 1 &  2 \\ 
$ \bar \nu   p \rightarrow l^+  K^-  p $     &     -F  & 2 &$-\frac{1}{2}$  &1 & -2 & -1 & 1 & 1 \\ 
$ \bar \nu   p \rightarrow l^+ \bar K^0  n $ &   -D-F & 1 & $\frac{1}{2}$ & 1 & -1 & -2 & 0 & -1 \\ \hline \hline
 \end{tabular}
\caption{Constant factors  appearing in the hadronic current}\label{tb:currents}
 \end{center}
\end{table*}


\begin{thebibliography}{}


\bibitem{Boyd:2009zz}
  S.~Boyd, S.~Dytman, E.~Hernandez, J.~Sobczyk and R.~Tacik,
  AIP Conf.\ Proc.\  {\bf 1189} (2009) 60.

\bibitem{Leitner:2006ww}
  T.~Leitner, L.~Alvarez-Ruso and U.~Mosel,
  Phys.\ Rev.\  C {\bf 73} (2006) 065502.

\bibitem{Leitner:2006sp}
  T.~Leitner, L.~Alvarez-Ruso and U.~Mosel,
  Phys.\ Rev.\  C {\bf 74} (2006) 065502.

\bibitem{Benhar:2010nx}
  O.~Benhar, P.~Coletti and D.~Meloni,
  Phys.\ Rev.\ Lett.\  {\bf 105} (2010) 132301.

\bibitem{Martini:2010ex}
  M.~Martini, M.~Ericson, G.~Chanfray and J.~Marteau,
  Phys.\ Rev.\  C {\bf 81}, 045502 (2010).

\bibitem{Amaro:2010sd}
  J.~E.~Amaro, M.~B.~Barbaro, J.~A.~Caballero, T.~W.~Donnelly and C.~F.~Williamson,
  Phys.\ Lett.\  B {\bf 696} (2011) 151.

\bibitem{Nieves:2011pp}
  J.~Nieves, I.~Ruiz Simo and M.~J.~Vicente Vacas,
  Phys.\ Rev.\  C {\bf 83} (2011) 045501.



\bibitem{AlvarezRuso:1998hi}
  L.~Alvarez-Ruso, S.~K.~Singh and M.~J.~Vicente Vacas,
  Phys.\ Rev.\  C {\bf 59} (1999) 3386.

\bibitem{Sato:2003rq}
T.~Sato, D.~Uno, T.~S.~H. Lee, Phys. Rev. {\bf C67} (2003) 065201.

\bibitem{Graczyk:2009qm}
K.~M. Graczyk, D.~Kielczewska, P.~Przewlocki, J.~T. Sobczyk,
Phys. Rev. {\bf D80}
  (2009) 093001.

\bibitem{Hernandez:2007qq}
  E.~Hernandez, J.~Nieves and M.~Valverde,
  Phys.\ Rev.\  D {\bf 76} (2007) 033005.

\bibitem{Leitner:2008wx}
  T.~Leitner, O.~Buss, U.~Mosel and L.~Alvarez-Ruso,
  Phys.\ Rev.\  C {\bf 79} (2009) 038501.

\bibitem{Leitner:2010jv}
  T.~Leitner and U.~Mosel,
  Phys.\ Rev.\  C {\bf 82} (2010) 035503.


\bibitem{Hernandez:2010bx}
  E.~Hernandez, J.~Nieves, M.~Valverde and M.~J.~Vicente Vacas,
  Phys.\ Rev.\  D {\bf 81} (2010) 085046.


\bibitem{Lalakulich:2010ss}
  O.~Lalakulich, T.~Leitner, O.~Buss and U.~Mosel,
  Phys.\ Rev.\  D {\bf 82} (2010) 093001.

\bibitem{VicenteSingh}
  S.~K.~Singh and M.~J.~Vicente Vacas,
  Phys.\ Rev.\  D {\bf 74}, 053009 (2006).


\bibitem{Mintz:2007zz}
  S.~L.~Mintz and L.~Wen,
  Eur.\ Phys.\ J.\  A {\bf 33} (2007) 299.

\bibitem{Dewan}
  H.~K.~Dewan,
  Phys.\ Rev.\  D {\bf 24}, 2369 (1981).

%
\bibitem{Shrock}
  R.~E.~Shrock,
  Phys.\ Rev.\  D {\bf 12}, 2049 (1975).

\bibitem{Amer:1977fy}
  A.~A.~Amer,
  Phys.\ Rev.\  D {\bf 18}, 2290 (1978).


\bibitem{Mart:2009}
G.B. Adera, B.I.S. Van Der Ventel, D.D. van Niekerk and T. Mart,
Phys.\ Rev.\ C {\bf 82}, 025501 (2010). 

\bibitem{RafiAlam:2010kf}
  M.~Rafi Alam, I.~Ruiz Simo, M.~Sajjad Athar and M.~J.~Vicente Vacas,
  Phys.\ Rev.\  D {\bf 82}, 033001 (2010).

\bibitem{Hayato:2009zz}
  Y.~Hayato,
  Acta Phys.\ Polon.\  B {\bf 40} (2009) 2477.

\bibitem{Gallagher:2002sf}
  H.~Gallagher,
  Nucl.\ Phys.\ Proc.\ Suppl.\  {\bf 112} (2002) 188.

\bibitem{Casper:2002sd}
  D.~Casper,
  Nucl.\ Phys.\ Proc.\ Suppl.\  {\bf 112} (2002) 161.

\bibitem{Zeller:2003ey}
  G.~P.~Zeller,
  arXiv:hep-ex/0312061.

\bibitem{Andreopoulos:2009rq}
  C.~Andreopoulos {\it et al.},
  Nucl.\ Instrum.\ Meth.\  A {\bf 614} (2010) 87.


\bibitem{Solomey:2005rs}
  N.~Solomey  [Minerva Collaboration],
  Nucl.\ Phys.\ Proc.\ Suppl.\  {\bf 142}, 74 (2005).

\bibitem{Kobayashi:2005}T. Kobayashi  Nuclear Physics B (Proc. Suppl.) 143 (2005) 303.

\bibitem{Mezzetto:2010}  M. Lindroos and 
M. Mezzetto, {\it Beta Beams: Neutrino Beam}, (Imperial College, London, 2009).

\bibitem{Scherer:2002tk}
  S.~Scherer,
  Adv.\ Nucl.\ Phys.\  {\bf 27}, 277 (2003).

\bibitem{Cabibbo:2003cu}
  N.~Cabibbo, E.~C.~Swallow and R.~Winston,
  Ann.\ Rev.\ Nucl.\ Part.\ Sci.\  {\bf 53}, 39 (2003).

\bibitem{Oller:2006yh}
  J.~A.~Oller, M.~Verbeni and J.~Prades,
  JHEP {\bf 0609} (2006) 079.

\bibitem{Paschos:2005} O. Lalakulich and E. A. Paschos Phys. Rev. D 71, 074003 (2005).

\bibitem{Leitner:2008ue}
  T.~Leitner, O.~Buss, L.~Alvarez-Ruso and U.~Mosel,
  Phys.\ Rev.\  C {\bf 79}, 034601 (2009).

\bibitem{hep-ph/9211247} 
  M.~N.~Butler, M.~J.~Savage and R.~P.~Springer,
  Nucl.\ Phys.\ B\ {\bf 399}, 69  (1993).

\bibitem{arXiv:0907.0631} 
  L.~S.~Geng, J.~Martin Camalich and M.~J.~Vicente Vacas,
  Phys.\ Rev.\ D\ {\bf 80}, 034027  (2009).


\bibitem{Doring:2006ub}
  M.~Doring, E.~Oset, S.~Sarkar,
  Phys.\ Rev.\  {\bf C74}, 065204 (2006).

\bibitem{Taylor:2005zw}
  S.~Taylor {\it et al.} [ CLAS Collaboration ],
  Phys.\ Rev.\  {\bf C71}, 054609 (2005).

\bibitem{AguilarArevalo:2011sz}
  A.~A.~Aguilar-Arevalo, C.~E.~Anderson, S.~J.~Brice, B.~C.~Brown, L.~Bugel, J.~M.~Conrad, R.~Dharmapalan, Z.~Djurcic {\it et al.},
  [arXiv:1102.1964 [hep-ex]].


\bibitem{AguilarArevalo:2009ww}
  A.~A.~Aguilar-Arevalo {\it et al.} [MiniBooNE Collaboration],
  Phys.\ Rev.\  D {\bf 81}, 013005 (2010).

\bibitem{Honda:2006qj}
  M.~Honda, T.~Kajita, K.~Kasahara, S.~Midorikawa and T.~Sanuki,
  Phys.\ Rev.\  D {\bf 75} (2007) 043006.

\bibitem{Ashie:2005ik}
  Y.~Ashie {\it et al.}  [Super-Kamiokande Collaboration],
  Phys.\ Rev.\  D {\bf 71} (2005) 112005.

\bibitem{arXiv:1109.3262} 
  K.~Abe, T.~Abe, H.~Aihara, Y.~Fukuda, Y.~Hayato, K.~Huang, A.~K.~Ichikawa and M.~Ikeda {\it et al.},
  arXiv:1109.3262 [hep-ex].

\bibitem{minflux}
L. Loiacono, MINERvA-doc-3042-v1, 
at http://minerva-docdb.fnal.gov/.

\bibitem{Adjei:1981nw}
  S.~A.~Adjei, D.~A.~Dicus, V.~L.~Teplitz,
  Phys.\ Rev.\  {\bf D24}, 623 (1981).

\bibitem{Kitagaki:1986ct}
  T.~Kitagaki, H.~Yuta, S.~Tanaka, A.~Yamaguchi, K.~Abe, K.~Hasegawa, K.~Tamai, S.~Kunori {\it et al.},
  Phys.\ Rev.\  {\bf D34}, 2554-2565 (1986).


\end{thebibliography}
\end{document}